# BitAV: Fast Anti-Malware by Distributed Blockchain Consensus and Feedforward Scanning

Charles Noyes, *cnoyes@usc.edu*

***Synopsis*—In the age of information, of the Internet, the protection of our most vital infrastructure becomes increasingly important. Moores law continues to prove accurate, with the number of transistors on standard integrated circuits doubling about every two years, but virus scanning applications have not innovated on the same level and development has stagnated. Thus, the attack surfaces become larger and the targets more lucrative, while the defensive mechanisms are failing to improve at a comparable rate.**

**I present the design and implementation of a novel anti-malware environment called BitAV. BitAV allows for the decentralization of the update and maintenance mechanisms of the software, traditionally performed by a central host, and uses a staggered scanning mechanism in order to improve performance. The peer-to-peer network maintenance mechanism lowered the average update propagation speed by 500% and is far less susceptible to targeted denial-of-service attacks. The feedforward scanning mechanism significantly improved end-to-end performance of the malware matching system, to a degree of an average 14× increase, by decomposing the file matching process into efficient queries that operate in verifiably constant ($O(1)$) time.**

## I. Introduction

As the use of the Internet, and other massively networked systems like it, becomes increasingly widespread, the ease with which viruses proliferate grows with it. The result is the need for technologies designed to block these viruses, generically called malware (malicious software), at all major network stops, but especially at the terminal end-user point. The throughput of most end-user's network connections, and thus the amount of potential data consumption, is greatly increasing as well. While network-based detection systems have reached speeds of over 1Gb/s, the speed of actual virus scanning and malware preventive systems has not kept pace.

The amount of new malware released onto the public Internet is exploding [1]. As most anti-virus software currently filters suspect files through string matching against pseudo-unique identifiers, each new malware sample, and variant sub-sample, requires its own signature [2]. Thus the size of a anti-virus's signature set $S$ is related to $\gamma$, the number of all known malware samples, $S \propto \gamma$. Please do note that 'signature' is used interchangeably with 'identifier,' both of which mean the unique value resulting from the inputting of the candidate malware sample into a cryptographic hash function (a function which maps an arbitrary input to a set-length output with uniform distribution).

Because virtually all anti-malware programs devote most of their resources to the matching of these signatures $S$ to some arbitrary input stream, usually with an exact matching algorithm, the two main factors that determine the effectiveness of a solution are the ratio of detected to undetected inputs (possibly taking into account the rate of false-positives, although this is only a problem in solutions that utilize regular-expression-based multi-pattern algorithms) and the scalability of the signature set. The second factor is due to the possibility of seemingly highly efficient implementations that are really only efficient in-memory, as they rely heavily on constant (very expensive) disk accesses, and therefore degrade user experience elsewhere in the system.

Obviously all of these possible pattern matching schemes rely on having a known pattern set that acts as the corpus that inputs are matched against. In this case, it is the known malware identifiers. Thus, any anti-virus solution that aims to protect its users from future malware types and variants must have update mechanisms in place which are able to update the known pattern set. The apparent solution is just to have a centralized update server, but this is sub-optimal, especially for open-source efforts, because of the cost associated with it and the fact that it acts as an obvious and openly-facing target for malicious attackers.

An ideal anti-malware system would be wholly efficient and extremely fast, but the two are gen-erally at odds with one another, and thus trade-offs must be made in the search for an acceptable middle-ground. Really the aim of this project is to find that middle-ground.

Objective evaluation shows that our solution, BitAV, is an effective architecture that is more optimal than any currently available commercial or researched/published solution. Specifically, we show:

- **Fast scanning speed with less memory usage:** By layering a cache-efficient bloom filter on top of the more costly bloomier filter, BitAV manages to increase end-to-end throughput of the average-case input by $14\times$, and requires less memory to do so than traditional algorithms.
- **Scalability:** BitAV can handle large numbers of signatures with ease, and further space-efficiency improvements in order-and-match construction within our data structures will further improve scalability.
- **Decentralized updates and maintenance:** The community of users using and maintaining BitAV are provided a trustable, timeless conduit by which to work together that is not dependent on any centralized authority or schema other than cryptographic verification. This is accomplished through the use of a novel blockchain variant.
- **Easily implementable on all types of devices:** BitAV should work on any architecture, provided that it has enough RAM and disk space to store the identifiers and load them into the memory. The low memory and disk space usage contributes to this.

### A. Virus Scanning Techniques Overview

**Signature Matching:** checking if a file is a known virus, or contains bytecode known to be malicious, by searching for the hash of the identifier.

**Heuristic Analysis:** testing for polymorphism by executing the virus and searching for known malicious identifiers in-memory.

**Behavioral Analysis:** checking if a file contains a completely unknown virus by running the file in an emulated environment; the downside is the large overhead of the emulation.

While this paper explores novel methods to apply the first type of scanning mechanism, signature matching, the others should not be discounted. In the future, when computational resources are so vast as to render their cost moot, they will likely be the best choice. They are not currently because of the associated reduction in end-to-end speed. Thus, this paper really focuses on the creation of a bridge between the currently used, and increasingly outdated, method and the likely future of anti-malware. The fostered debate around this transitional period [3], [4], is ongoing, but the consensus view seems to be that, for the time being, signature matching is the best option.

There are two main approaches to signature matching: exact and rolling. Exact signatures are of whole files, whereas rolling signatures 'roll' over some sections $n$ of the file $F$, such that $n = F/k$ where $k$ is the size of each section. Rolling signatures are useful because if even one bit is flipped in a malware sample that codes for an identifier, the corresponding exact signature will change dramatically (as a result of the waterfall effect of hash function computation), while the rolling signatures will remain mostly the same (perhaps only one section will be divergent) [5]. Thus, exact signatures are more precise, whereas the rolling signatures are more likely to detect slightly modified variants.

## II. Methods/Design

This paper centers around the design and implementation of 'BitAV,' an anti-malware system that uses novel techniques to propagate malware identifiers along a network of users in conjunction with an extremely efficient pattern-matching scheme of my own design to create the optimal anti-malware solution. The scanning mechanism utilizes a bloom filter [6] and one of its derivative data structures, the bloomier filter [7], to create a structure that allows for constant time key-value queries, without the high probability of false-positives that comes with probabilistic data stores.

This is not the first use of bloom filters to speed up pattern matching ( [8], [9]), nor the first to use tiered look-up systems based on stratified bloom filter layering ( [10], [11]). This is, however, the first implementation that takes advantage of cheap hash functions, feedforward logical flow, and cache-resident (or in the case of systems that run a dedicated processor, texture memory-resident and massively parallelized) architecture. This module is a more final refinement of previous works, namely [5] and [10].

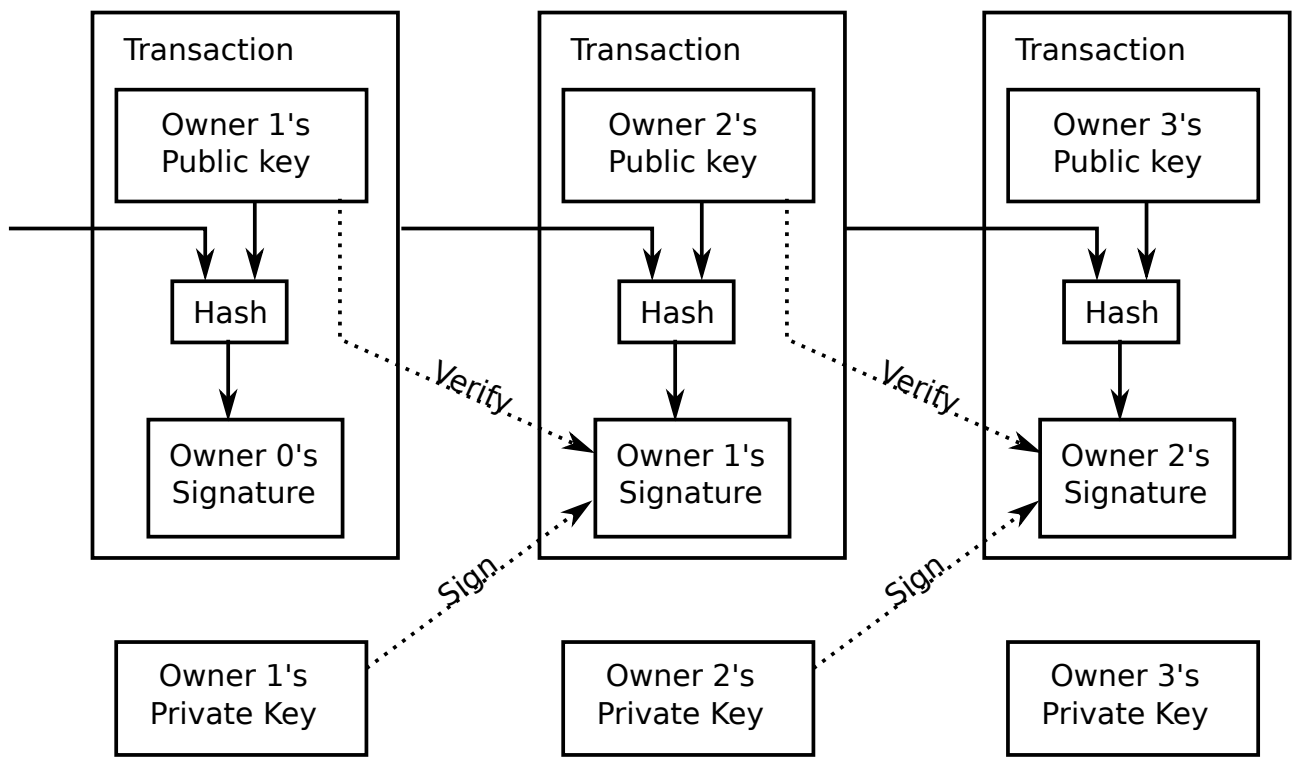

Fig. 1: Currency Blockchain [15]

The networking module acts as a way for users across a decentralized [12] network to both receive and transmit information trustlessly, and works to improve the reliability and efficiency of the update network. It does this using blockchain architecture that allows for distributed anonymous consensus among peers, with the 'vote' (an abstracted representation of influence) weighting being a result of computational power expended. Thus, in order to overpower the rest of the network, a user (or a group of malicious actors) would need $> 50\%$ network power to gain control [13]. This solves the Byzantine Generals' problem nicely, as the expenditure of computational power scales directly with the amount of capital required to obtain it (either through hardware or, more often, electricity costs) [14]. This network model is very similar, and is actually derived from, the core Bitcoin protocol, laid out by the pseudonymous Satoshi Nakamoto in [15].

*A. Blockchain Architecture Overview*

Originally created by the visionary Satoshi Nakamoto, Bitcoin has revolutionized the business of digital currencies [16]. Bitcoin is, however, only one of the innumerable number of potential applications of the blockchain (illustrated in Fig. 1). It has the potential to completely decentralize data storage, reputation systems, even democratic voting. These are all done through the creation of self-executing digital contracts that are backed by intelligent assets (cryptocurrency 'coins'). Because there is monetary value associated with these contracts, there is an incentive to make sure that they are 'correct,' and this (coupled with the proof-of-work [15] system present in these applications) allows for the coordination of networks that control valuable information over an anonymous network (in this case the information is the transactions that determine currency ownership). Prior the invention of these mechanisms, it was simply not possible to coordinate large numbers of individual activities into a cohesive network without a centralized governing body to watch over and verify the proceedings [17]. This problem of coordination is a well-known problem in the field of distributed networking, originally outlined in the 80's [18] and more recently encapsulated by the umbrella 'Byzantine Generals Problem' [19], alluded to earlier.

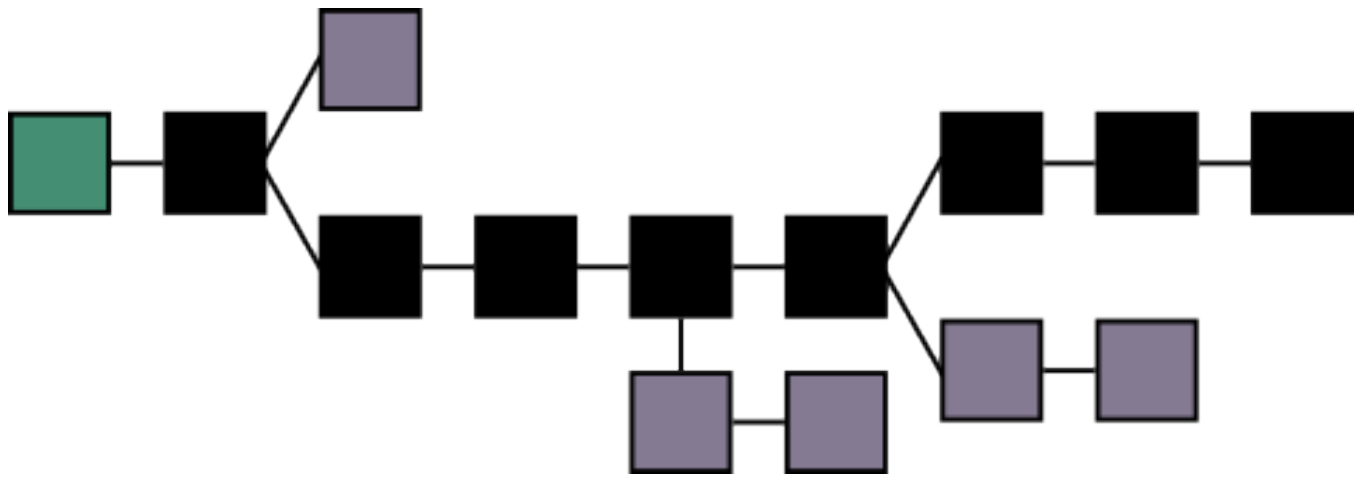

Fig. 2: Orphan Chain Competition

The Generals problem questions how individual computer systems can come to a consensus without a method of omniscient verification (which a central body would provide), in such a way that the network is resilient to attacks by bad actors[1]. It posits that three divisions of the Byzantine army are camped outside an enemy city in hopes of conquering it; an independent commander directs each division and, in order to be successful, all three must attack at the same time [19]. The generals can only communicate through an unreliable messenger, which may be influenced by a traitor in the group who is actively trying to derail the generals' efforts.

A blockchain solves this problem by forcing transparency among the groups using it, and uses cryptographic measures to allow for independent verification of transmitted information by all groups in the network. In order to send a new message (or 'mine' a new block), significant computational power must be expended; this makes it both expensive and cumbersome for a bad actor to coordinate an attack against the network. Blockchain protocols thus ensure that transactions (included in each block, and the truly 'valuable' part of the system) are valid and secure, as long as $> 50\%$ of the network is non-malicious [14], [16], [20]. The possibility of multiple competing chains being extended at once, illustrated in Fig. 2, is addressed by having each individual select the longest chain they know of; as long as the proof-of-work computations are accurate, the longest chain is verifiably the most difficult to compute, and therefore the least likely to have been compromised or be a false chain.

[1]"The Byzantine Generals Problem seems deceptively simple. Its difficulty is indicated by the surprising fact that if the generals can send only oral messages, then no solution will work unless more than two-thirds of the generals are loyal. In particular, with only three generals, no solution can work in the presence of a single traitor." [19]

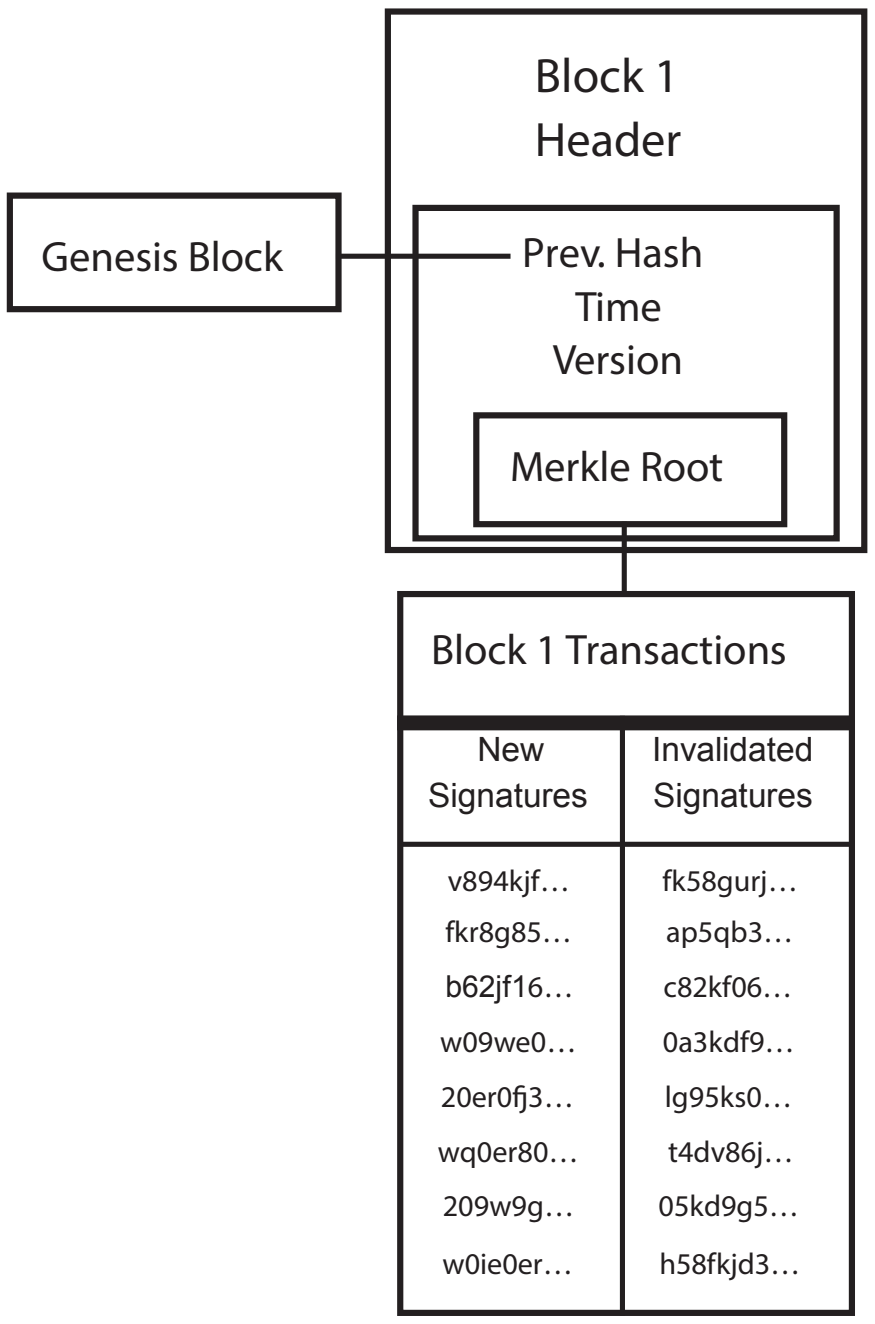

Fig. 3: BitAV Blockchain

### *B. BitAV Blockchain*

The BitAV blockchain is similar to the core Bitcoin implementation's, in that it uses cryptographic proofs to extend and verify the chain, but it radically differs in the way that the transaction field works. Traditionally, the transaction ('tx') field houses a forward-flowing narrative of all known state-changes for the currency it is recording [21]. BitAV replaces the tx field with two information stores: the identifier and invalidation fields. The identifier field allows for the addition of novel malware identifiers; this does not mean 'new' signatures, just those not currently present on the chain in some other block. This is illustrated in Fig. 3.

Each field in the block header (essentially the meta-data section) is needed to ensure verifiability and consensus. First, the version number is needed to prevent errors resulting from hard forks (updates that would break compatibility with previous versions). The time is the approximate creation time of the block and must be within the calculated acceptable range for each new block; BitAV uses the timestamp both to recalculate difficulty (for use in the proof-of-work verification ['mining' in Bitcoin's terminology]) and in some checks that work to ensure chronological sanity and canonicity. The previous hash field is what really makes the blockchain a 'chain,' as it necessitates the inclusion of the most recent block's hash in the next block to be created; because of the difficulty of reverse-engineering the has function used (SHA256, in this case), the longer the chain the more difficult it is to recreate a verifiable blockchain. For reference, the difficulty in recreating the current Bitcoin chain is approximately $\frac{1}{(2^{256})^n}$, where $n$ is the current blockchain length. Note that this is the worst-case time when attempting to recreate the exact hashes of all current blocks using falsified transactions; individual blocks would reduce it to $\frac{1}{(2^{256})^{n-k}}$, where $k$ is the depth of the recreation attempt from the 'top' of the chain. Additionally, this is assuming SHA-256 is used as a hash function; should another hash function of digest length $d$ be used, the approximate odds of finding a collision are $\frac{1}{2^d}$. Finally, not pictured is the 'nonce' that is included in the header so that the hash of said header can be quickly modified by changing the nonce. Users extend the chain by finding a nonce that results in the hash of the blocks header being a lower value than the calculated difficulty level [15].

This architecture can actually be used for any networks whose aim is to share inherently valuable information across a network of users. The only stipulation is that the information have some unique characteristics. In this case, BitAV peers working to extend the chain can leverage open-source databases to check whether a submitted identifier is known to be good or bad; in either case, it allows for some level of pre-screening. There is still the possibility of attack by the submission of identifiers that are currently unknown to be good or bad, or more generally the submission of universally unknown identifiers for some malicious purpose.

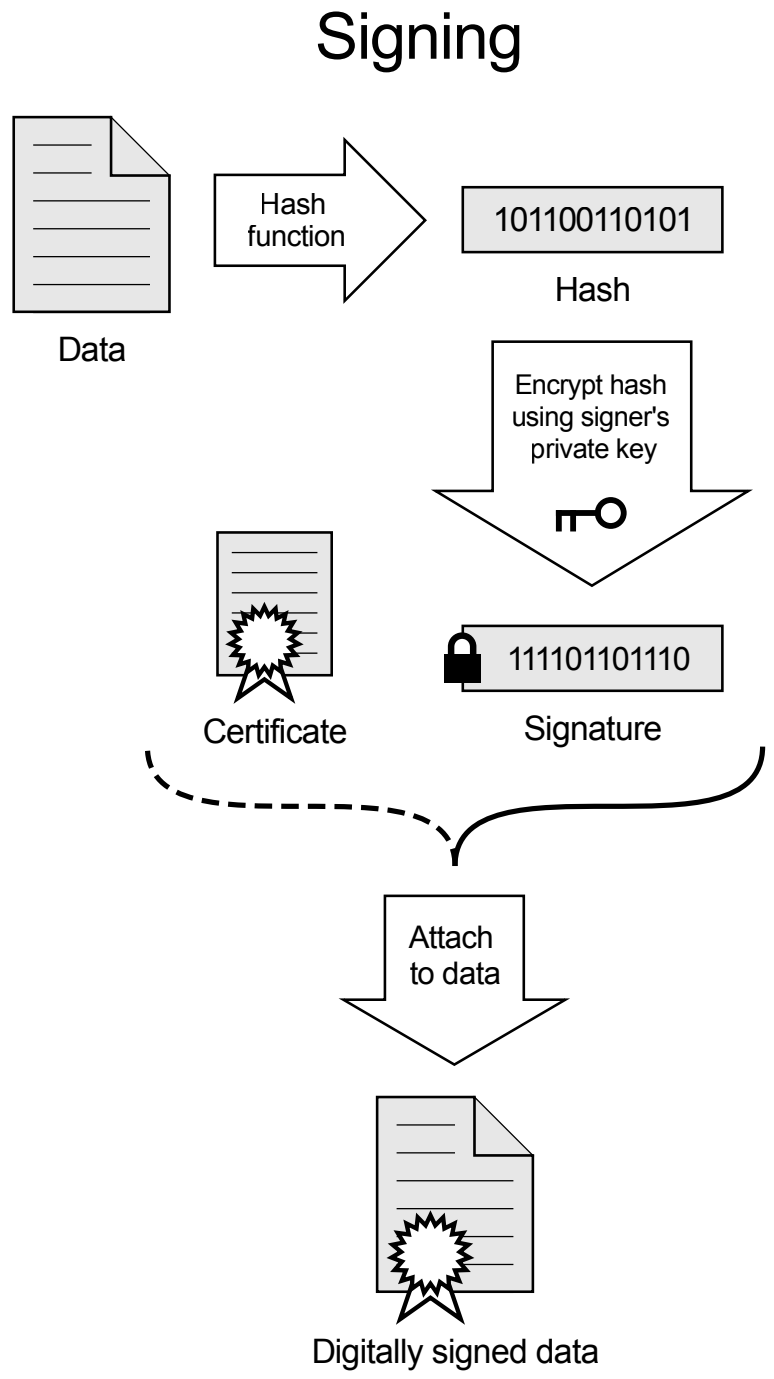

Fig. 4: Cryptographic Signing of Data [22]

This vector of attack is mitigated, almost in its entirety, by a novel voting scheme that we call 'minority transaction consensus.' This is perhaps the most innovative feature of our network's design, and it is what drives the 'invalidation' field of our blockchain. To fully appreciate the significance of this scheme one must understand why it is that this would never work on a currency network, such as Bitcoin (we theorize this is the reason that no other organization has though to implement such a protocol). When dealing with immutable currencies, the ability to just 'rewind' transactions and return currency to an entity that had appeared to have spent it is not allowable, as it opens the door for far too much fraud and actually would turn Bitcoin into an even more abusable version of credit card chargebacks (the preclusion of which Bitcoin touts as one of its greatest strengths). In our network, however, each signature is not actually valuable to an individual, but to the network as a whole in a more probabilistic sense. Allowing individuals to reclaim currency which they have appeared to have spent, and is accepted into the network as canon and shown to have been validated to all peers, is harmful on a case-by-case level, whereas the possibility of $0.5\%$ of all malware identifier invalidations being malicious would only marginally effect the network *as a whole*.

We understand that given the newness of the blockchain architecture and the nature of these problems being less scientific and more game-theory oriented, the authors encourage all interested parties to read up more fully on these concepts in [13]–[15], and especially [17], which is far more approachable than many of the more highly specialized studies.

Each new identifier submission holds the general format of:

`[Identifier] [Pubkey] [Signature]`

The signature is the cryptographically signed hash of the identifier, and to verify that the submission is not a forgery. Verification can take place by using the included public key to decrypt the signature, hashing the decrypted value, and then checking the hash of the identifier against the digest of the decrypted value's hash [23], [24]. It is a very simple and well established procedure, and the signature step is illustrated in Fig. 4.

Because we are able to ensure that all of the submissions under a specific key-pair are, in fact, generated by using that key-pair to sign the submission, we can 'track' a user through the network by search from all the occurrences of their public key in the 'new identifiers' transaction field of each block on the chain. Knowing this, we posited that a user's relative 'trust' could be evaluated by counting these occurrences and using the number of times they have altruistically added value to the network (in the form of new identifier submissions) as the weighting factor in deciding their trust.

When users trying to extend the chain are actively broadcasting their status as current miners, they receive both requests to add new identifiers and 'votes' from users attempting identifiers they believe are invalid. Invalidations are only possible for signatures added within the last 10 (note that this is an arbitrary limit and can be easily modulated once large-scale testing is done) blocks, so that the backbone of the blockchain is unmanageably canonical. A nice side effect of using new blocks to change the state of identifiers in older blocks is that the data within those older blocks is never changed (and thus the hash remains constant), so the prev. hash field is not compromised.

The actual calculation done by a miner in deter-

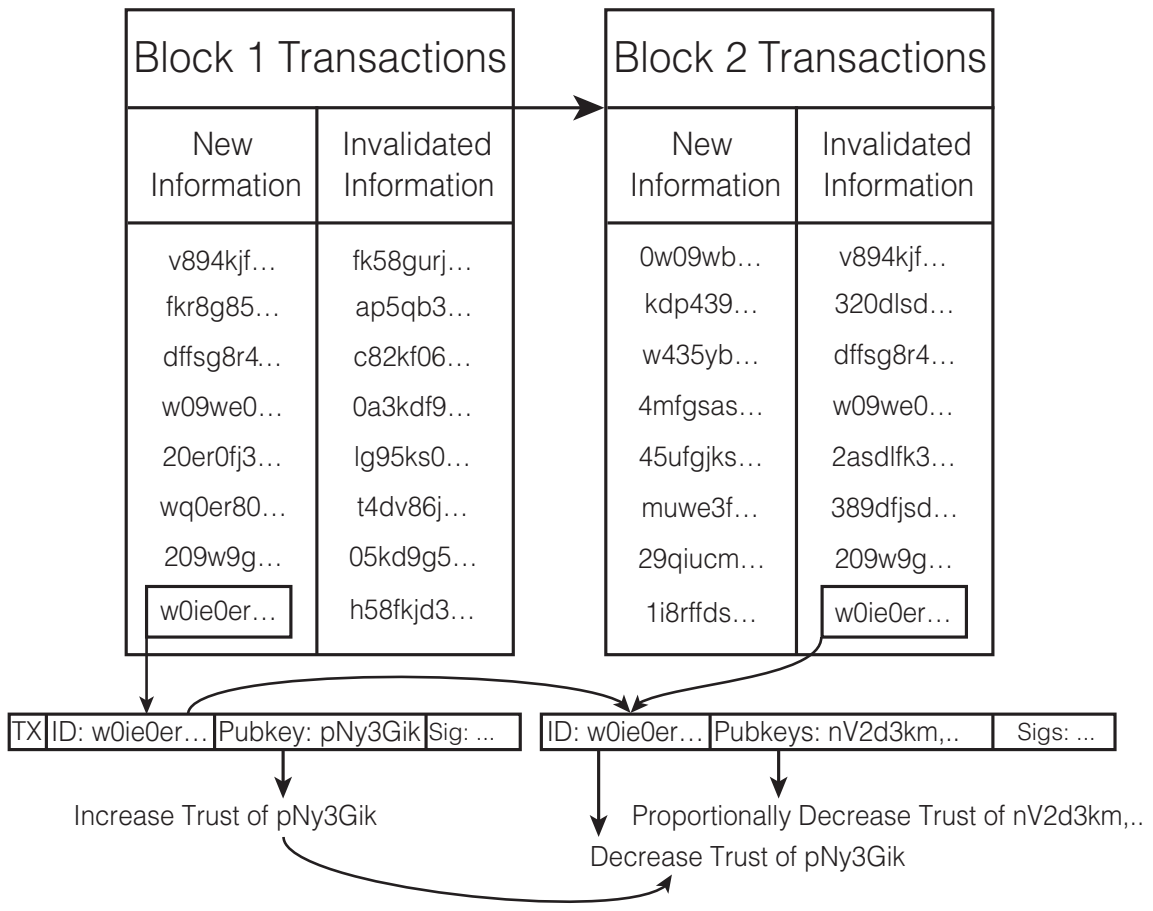

Fig. 5: BitAV Trust Determination Scheme

mining if an invalidation is allowable is simply:

$$\sum \text{Trust of Invalidators} - \sum \text{Trust of Submitter} \tag{1}$$

If the result of this is $> 0$, the invalidation goes through. There is no punishment for a failed invalidation vote because it would bloat the chain. If the outcome is of an invalidation, the submitter loses trust equal to that of the result of the subtraction, and the invalidators lose trust proportionally to their own trust level. For example, if an invalidation with 10 voters 10 trust total, and they have equal initial trust, they will all lose one 'trust.' However, if one has one more trust initially, they will lose more than the rest. While it may not seem intuitively 'fair' to give penalties for invalidations to the voters, it is necessary to ensure that a bad actor who has accrued lots of trust does not take over the network by invalidating the submissions of any opposers. Note that the trust points here are simplified for the example and are abstracted from the real implementation; never the less, the core ideas hold true.

Bit-AV exploits the knowledge that most input files are not malicious by using probabilistic data structures with exactly precise no-match accuracy to greatly speed up the process of file scanning. It first constructs a counting bloom filter [6] (a probabilistic data structure that uses bit vectors to efficiently map data [25]), illustrated in Fig. 6 from the set of known identifiers $S$. We start with an integer vector (as we want to allow for deletions [26]) of size $m$ buckets, all of which are set to $0$ during creation. For each identifier, $k$ hash functions are applied to its each signature $\alpha \in S$, resulting in hash digests $h_1(\alpha), h_2(h_1(\alpha)+\alpha), \ldots h_k((h_{k-1}(\alpha))+\alpha)$. Note the use of the result of the first hash digest in the second, and the second in the third, etc., as this allows us to use just one hash function while still being able to efficiently modulate $k$. To ensure the digests $h_1(\alpha), \ldots h_k((h_{k-1}(\alpha))+\alpha)$ are within the bounds of $1, \ldots m$, the result of each digest is modulated by $m$.

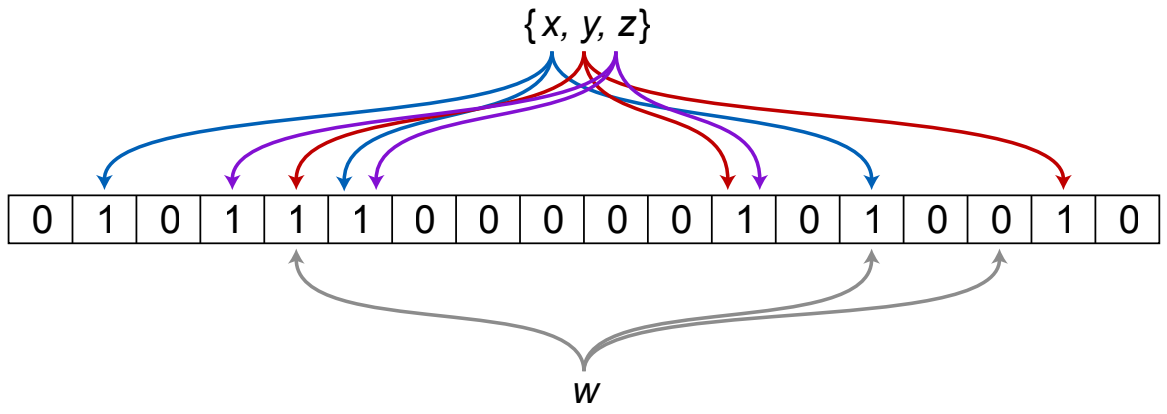

Fig. 6: Bloom Filter [9]

### C. *BitAV Scanning Mechanism*

Because of the possibility of hash collisions, in which two inputs result in the same output digest, we must consider the possibility of false positives. Given that after inserting $n$ keys into our table of size $m$, the probability of a specific bit being 0 is exactly

$$(1 - \frac{1}{m})^{kn}, \tag{2}$$

the probability of a false positive p is exactly

$$p = (1 - (1 - \frac{1}{m})^{kn})^k \approx (1 - e^{\frac{-kn}{m}})^k. \tag{3}$$

Finally, we can derive that given a target false probability $p$, the minimum value of k that will produce this probability is:

$$k = \frac{m}{n} \ln 2. \tag{4}$$

After construction of this bloom filter, we are left with a basic probabilistic data structure that can perform lookups in constant time while residing in the level 2 (L2) CPU cache. This filter will never produce a false-negative result, assuming proper implementation, but each lookup has the probability of being falsely positive $p$.

In a bloom filter's worst case scenario, the entire known set of data must be looked up in the filter to confirm an uncertain match from the hashing operation; this scenario can occur when a filter is too small for the number of elements it contains (and thus the false-positive rate would be extremely high). Hash collisions that result even before the table size modulation operation can occur, but they are far less likely.

Bloomier filters [7] solve this problem well in this case, as they allow for key-value lookups using a vector structure similar to those in bloom filters. The query runs in constant time and the space requirement is only $O(n\tau)$, where $\tau$ is the size of the bucket used in the order-and-match finder operation [11], [27].

$$Info = \bigoplus_{i=1}^{k} knownPatterns(Hash_i(Suspect)) \tag{5}$$

Equation 5 denotes the method by which information is returned from the bloomier structure. An n-ary XOR operation (symbolized by the $\bigoplus$ operator) is performed on the hash digests, 1 to $k$, of the suspect string(s). The digests $h_1(\alpha), \ldots h_k(\alpha)$ are used as indices in the bloomier index table (for example, the first position would be at $h_1(\alpha)$). The data in the index table at all of these positions, collectively, is XORed with the total index table, returning the information.

Because of the ease and efficiency with which Eq. 5 performs simple key-value lookups, and the usage of index table XORing, many parallels can be draw between our bloomier filters and structures like IBLT's [28], KBF's [29], and other bloomier-like data stores. While it is possible that one of these similar data structures outperforms a bloomier filter, our model can be easily adapted should that prove to be the case. As of now no clear consensus exists on the matter.

The integration of these two filtering methods naturally leads to a system in which information flows downward, stopping when it hits an impassible junction, but sometimes slipping through the cracks in the form of false-positives. I call this mechanism a feedforward bloom-bloomier filter (FBBF) because of the preclusion of a feedback loop developing and the necessitation of downward flow in the data stream. The only modifications come from the input disturbances caused by the networking module.

The full FBBF mechanism is illustrated in Fig. 7. The general format of the scanning algorithm is derived from [10], which uses a two-tiered bloom filtering mechanism. Unlike in our filter, their last step was a full pattern matching against a sub-set of the larger signature set (that was still fairly substantial). Our mechanism doesn't require this step, as it operates in constant time throughout.

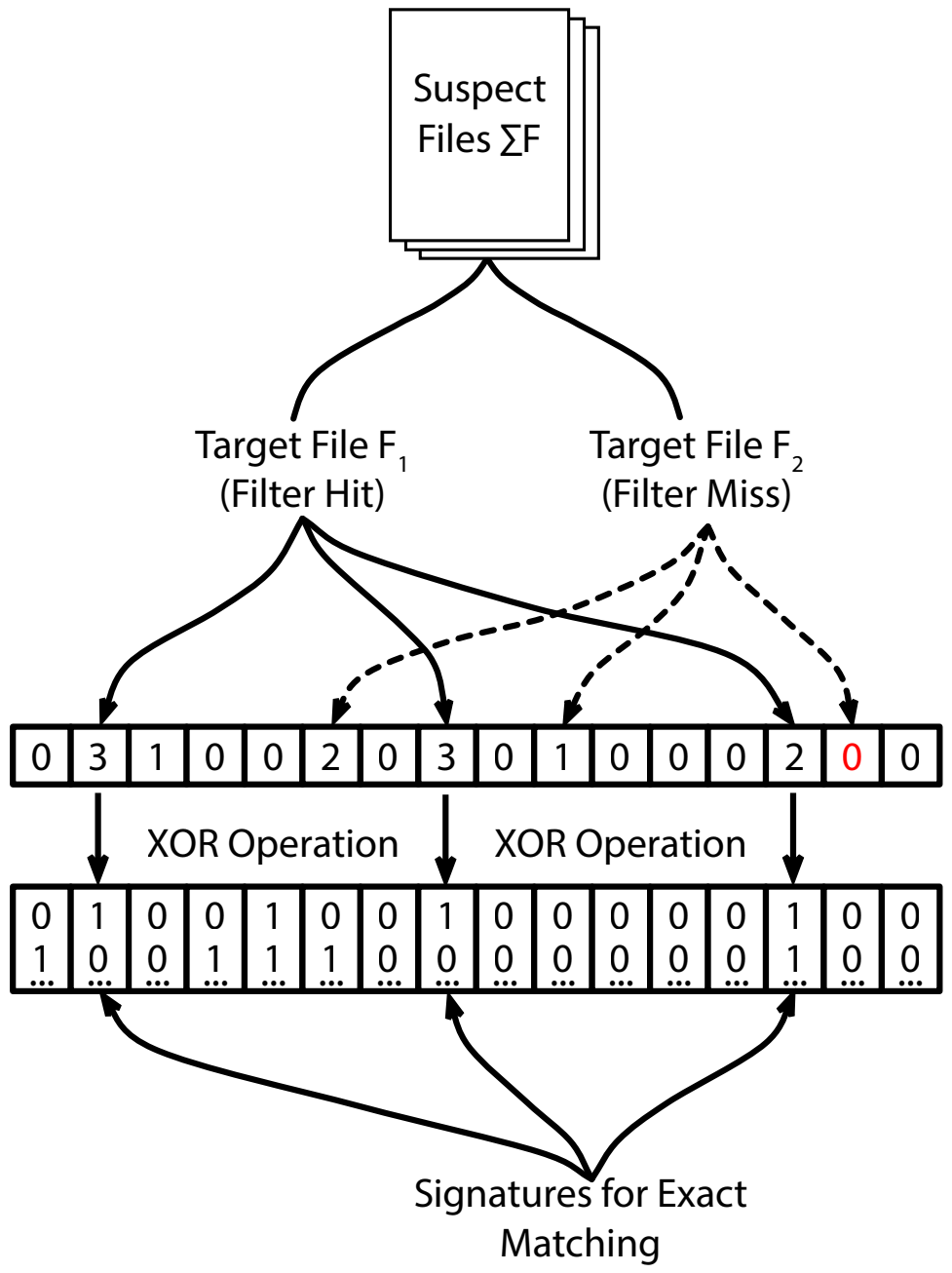

Fig. 7: Feedforward Bloom-Bloomier Filter

**BitAV-Hash**$(\sum) \rightarrow \delta$ takes the set of signatures $\sum$ and outputs an integer vector $\delta$ that contains the hash digests (signatures) of all known malware.

**BitAV-Screen**$(\delta, F) \rightarrow (\lambda, F_{suspect})$ constructs a feedforward filtering mechanism from $\delta$ and a bloomier filter. Each file $f \in F$ is scanned using $\phi$. The tuple ($\lambda$, $F_{suspect}$), where $F_{suspect} \subseteq F$, is the list of files matched by $\delta$, and $\lambda$ is string that the indices of the signatures actually matched in $F_{suspect}$.

**BitAV-HitScan**$(\lambda, \sum) \rightarrow \sum'$ takes $\lambda$ and outputs the set of signatures $\sum' \subseteq \sum$ that were matched during BitAV-Screen by querying our bloomier filter.

**BitAV-HitMatch**$(\sum', F_{suspect}) \rightarrow F_{malware}$ takes in the set of signatures $\sum'$, the set of files $F_{suspect}$, and outputs the set of files $F_{malware} \subseteq F_{suspect}$ matching $\sum' \bigoplus F_{suspect}$.

Note that the entire operation, including the final string matching, runs in constant time. The final 'exact matching' stage does not actually utilize an exact matching algorithm, as they run in linear time and would reduce the worst-case run to $O(n)$, but actually just XOR's the hash of the suspect file with the bloomier output. If the result is not 0 (as the XOR of two exactly similar objects is always 0), then the result was a false positive. Otherwise, the file is recognized as malicious and deleted.

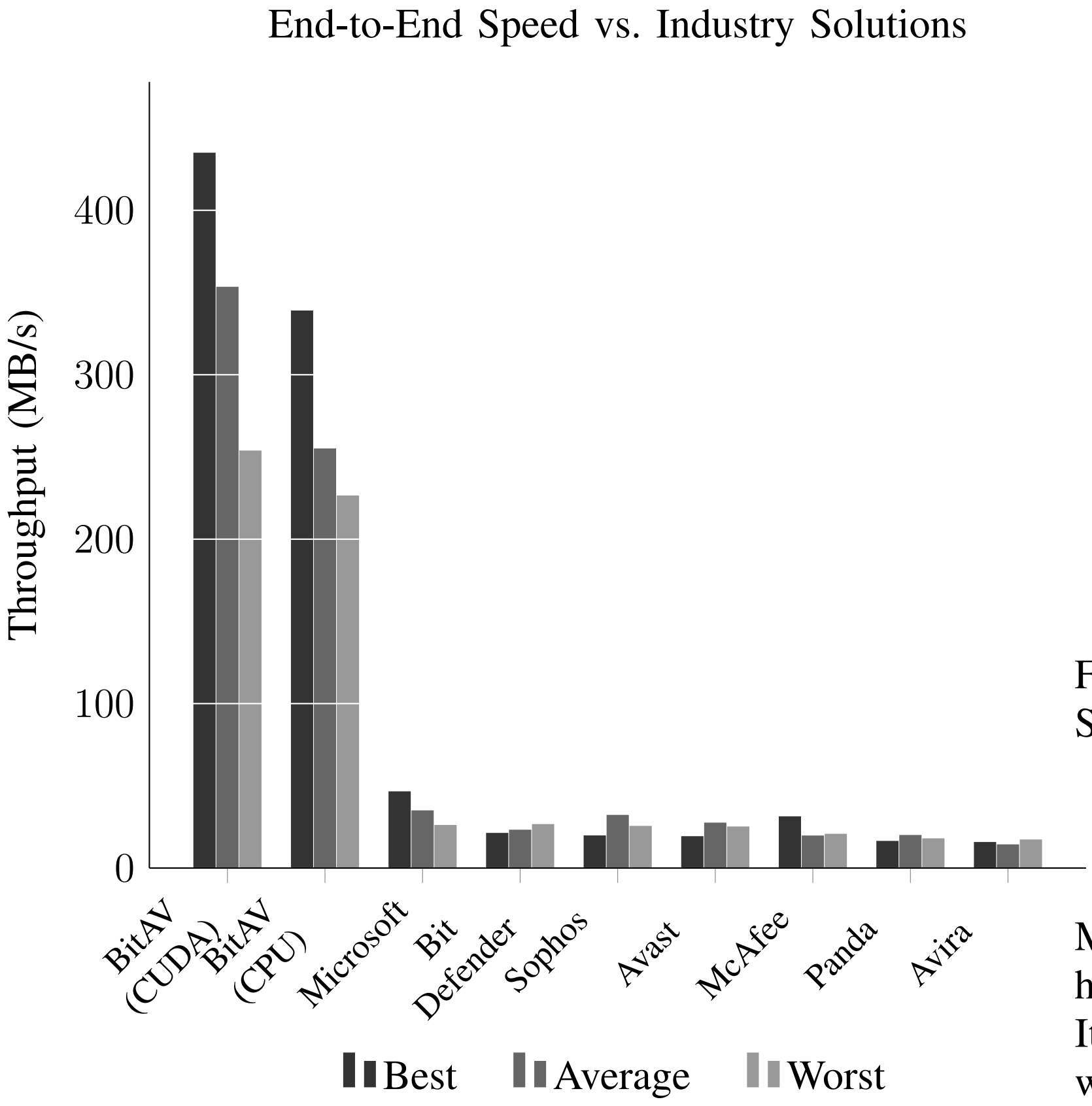

Fig. 8: Throughput Graph

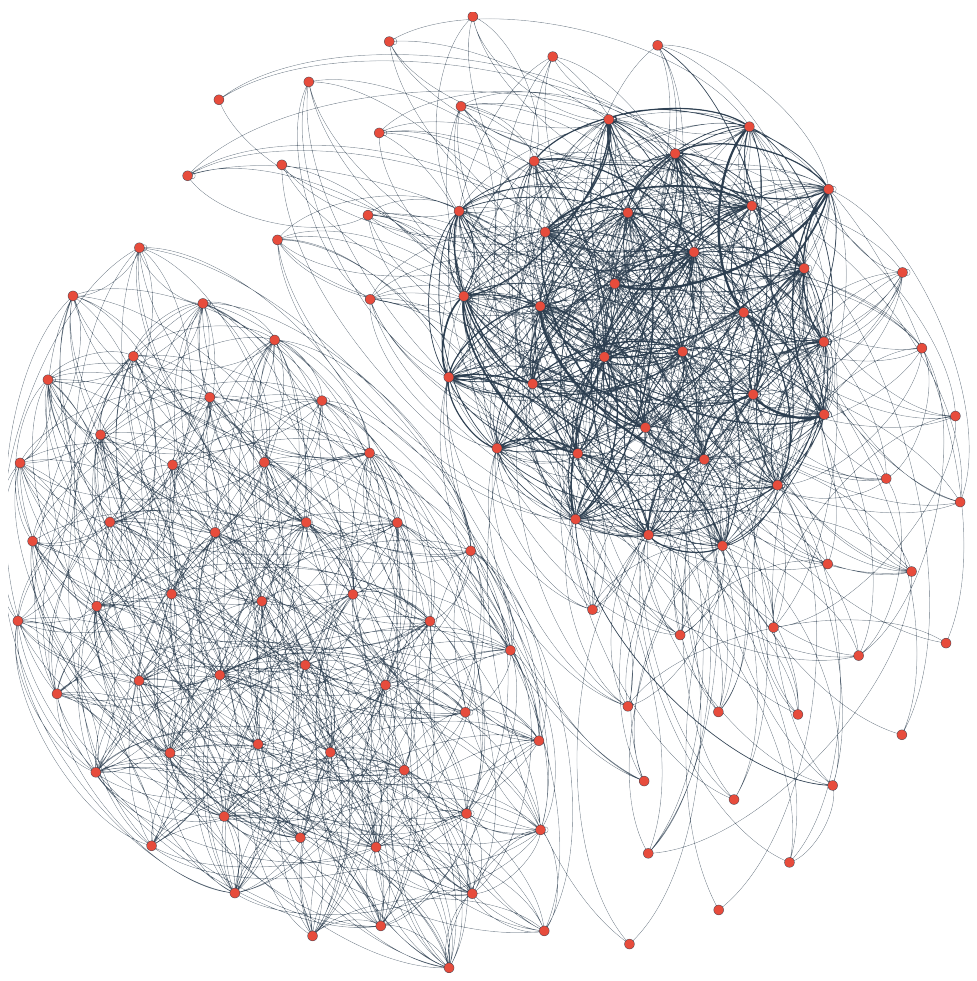

Fig. 9: Social Network Analysis of Real-World Simulation

## III. Results

My anti-virus implementation was, on average, 1,400% faster at the process of scanning files than the mean of the industry solutions tested against my implementation. This is including test data sets that were largely cache-misses or not already in the set of known malware for BitAV. I considered my average speed to be the mean of the 'average case' end-to-end speed tests ( 70% clean files, 30% malware binaries), with all of the end-to-end throughput test results represented in Fig. 8. Eliminating those cases brings the average speed increase to 8,500%, twice the average performance gain of [5] without optimizations for cache residency.

One of the main improvements on my original design was the parallelization of the signature generation for cases in which bulk file buffers were passed to BitAV. This was done through a simple CUDA MD5 implementation that was significantly faster at hashing large buffers than CPU-driven mechanisms. It was not, however, faster when the input stream was inconsistent (which would translate to on-access in the real world). Disk I/O speeds are increasing fast enough that within the next decade they will most likely not be the limiting factor for these kinds of pattern matching schemes.

Measurements of propagation time were much more difficult to obtain than scanning speed. To measure it, a network of servers (mostly AWS micro instances or virtual machines in accessible datacenters) was constructed to run my software. I set up a honeypot server on my own remote machine, and piped all of the garnered binaries into a test blockchain (which was shared with all of the other servers). I measured the difference between VirusTotal's first sample seen history and the time it took me to identify the sample (all of which I had to manually check to make sure they were, in fact, malicious); this yielded BitAVs data, and the data of the industry solutions was done by periodically checking VirusTotal's report API to monitor the detection status. The reason that most of my results are clustered into tiers is because VirusTotal's database of provider results for each sample is only updated periodically, leading to highly clustered results. Once again BitAV ran significantly faster than the tested industry solutions, 500% on average, and our results were even more statistically significant.

The last measurement I took was of the connections made between test network servers. The relationships are represented in a social network analysis graph in figure 4 (note that the OpenOrd ranking algorithm was used to arrange the nodes). The graph clearly shows that the nodes naturally clustered into a P2P-structured network. There are quite a few advantages to a networks hosting and maintenance being decentralized, namely the preclusion of denial-of-service attacks (assuming the protocol is not inherently vulnerable),

## IV. Conclusion

My findings showed that the implementation of the proposed design exceeded expectations in all areas of performance. When fully optimized for cache residency and with manually tuned bytecode our anti-virus was able to scan at a speed of over 350MB/s, making it a viable solution for network based scanning. Yes, it is very possible that an entire network could be secured through deep packet inspection by way of BitAV's revolutionary scanning mechanisms. The propagation speed tests similarly showed BitAV's clear preeminence over 'industry standard' solutions.

As more devices are brought online it becomes increasingly important to make sure that all networks are able to be secured with relative ease and no other solution provides as much modularity as one that is hosted by a community of its own users. Adaptation is limited only by the enthusiasm of the network's users to contribute to development, and previous open-source software development efforts in similar areas have shown this to be the least likely limiting factor. Furthermore, with the advent of malware that is developed and spread by a bad actor with influence over the institutions built around cyber defense (e.g. NSA's REGIN and SIGINT malware campaigns, whose identifiers have still not been added to a majority of the commercially available solutions), it is advantageous to users to know that there is no bias associated with the addition of new malware identifiers.

To be clear, my blockchain-based consensus scheme is the only one in existence that works for unvalidatable data across an anonymous network. The potential applications of this architecture are innumerable. I believe that once atomic binding of currency networks to informatory networks becomes a reality (most likely before the year is out), my networking architecture will have the potential to disrupt every industry based around valuable, time-critical data.

## V. Acknowledgements

I would like to thank prof. Chi So, information security department at the Viterbi School of Engineering at the University of Southern California, for introducing me to the field. In addition I would like to thank Pei Cao, Ozgun Erdogan, Sungmin Cho, prof. David Brumley, and Sang Kil Cha for providing useful discussion and source code for their implementations of Hash-AV and SplitScreen, respectively. Finally, VirusTotal provided access to their private research A.P.I. to gather data for the propagation speed test and malware samples for the scanning speed tests.